\documentclass{vgtc}                  
\ifpdf
  \pdfoutput=1\relax                   
  \usepackage{graphicx}                
  \DeclareGraphicsExtensions{.pdf,.png,.jpg,.jpeg} 
\else
  \usepackage{graphicx}                
  \DeclareGraphicsExtensions{.eps}     
\fi%

\graphicspath{{figures/}{pictures/}{images/}{./}} 

\usepackage{microtype}                 
\PassOptionsToPackage{warn}{textcomp}  
\usepackage{textcomp}                  
\usepackage{mathptmx}                  
\usepackage{times}                     
\usepackage{cite}                      
\usepackage{tabu}                      
\usepackage{booktabs}                  

\usepackage{amsmath}
\usepackage{amsfonts}
\usepackage{amssymb}
\usepackage{graphicx}
\usepackage{xcolor}
\usepackage{hyperref}

\graphicspath{{Pictures/}{figures/}}

\newcommand{\field}[1]{\mathbb{#1}}
\newcommand{\R}{\field{R}}

\newcommand{\domain}{\mathcal{M}}
\newcommand{\range}{\R}

\newcommand{\simplex}{\sigma}

\newcommand{\Index}{\mathcal{I}}

\DeclareMathOperator*{\argmin}{argmin} 

\onlineid{1004}

\vgtccategory{Research}

\title{Autoencoder-Aided Visualization of Collections of Morse Complexes}

\author{Jixian Li\thanks{e-mail: \{jixianli,vanboxel,josh\}@arizona.edu}\\ %
        \parbox{1.7in}{\scriptsize \centering Department of Computer Science \\ University of Arizona} %
\and Danielle Van Boxel\footnotemark[1]\\ %
     \parbox{1.7in}{\scriptsize \centering Program in Applied Mathematics \\  University of Arizona} %
\and Joshua A. Levine\footnotemark[1]\\ %
     \parbox{1.7in}{\scriptsize \centering Department of Computer Science \\  University of Arizona}}

\abstract{Though analyzing a single scalar field using Morse complexes is well studied, there are few techniques for visualizing a collection of Morse complexes. We focus on analyses that are enabled by looking at a Morse complex as an embedded domain decomposition. Specifically, we target 2D scalar fields, and we encode the Morse complex through binary images of the boundaries of decomposition. Then we use image-based autoencoders to create a feature space for the Morse complexes. We apply additional dimensionality reduction methods to construct a scatterplot as a visual interface of the feature space. This allows us to investigate individual Morse complexes, as they relate to the collection, through interaction with the scatterplot. We demonstrate our approach using a synthetic data set, microscopy images, and time-varying vorticity magnitude fields of flow.  Through these, we show that our method can produce insights about structures within the collection of Morse complexes.
} 


\teaser{
  \centering
  \includegraphics[width=0.83\linewidth]{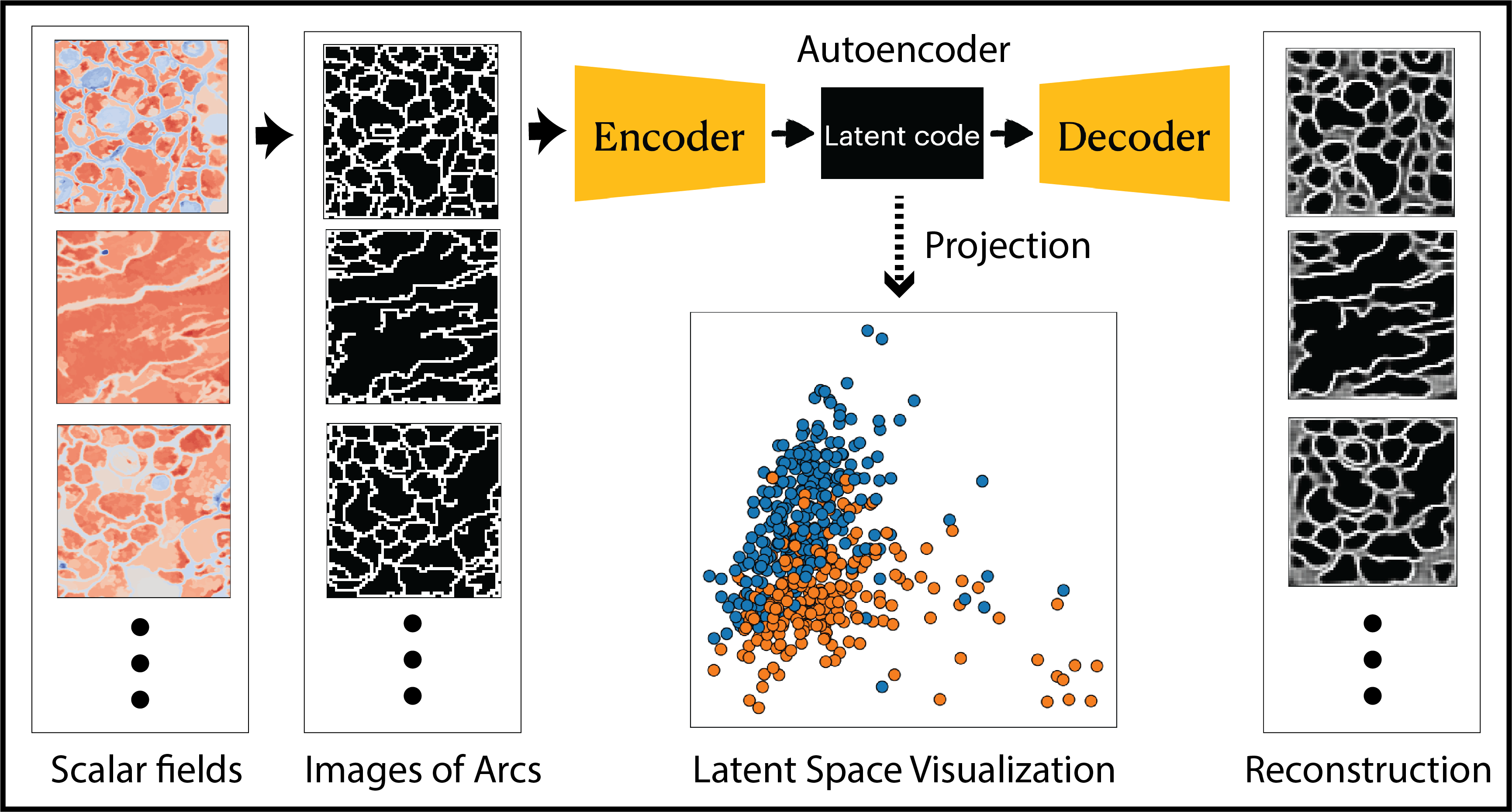}
  \caption{From a collection of scalar fields, we first compute the arcs of their Morse complexes. Then we convert the arcs of Morse complexes into image-based encoding to obtain a set of images of arcs. Next, we train an autoencoder to reconstruct those images. The autoencoder learns latent representations of those images of arcs. We use those latent representations as a proxy for the geometry of Morse complexes. Projecting those latent representations into a visual space helps us study the collection of Morse complexes.}
  \label{fig:teaser}
}

\begin{document}


\firstsection{Introduction}
\maketitle

Morse theory provides robust mathematical tools that can be used to study the topological structure of manifolds and functions defined on manifolds.
In particular, the Morse and Morse-Smale complexes partition a manifold based on the gradient flow behavior of a scalar function defined on the manifold.  
The resulting partitions serve as structural summaries of important features, and come imbued with a natural definition of scale.
Numerous works have demonstrated their utility in visualization and data analysis contexts across a wide variety of applications, including turbulent mixing~\cite{laney2006understanding}, porous materials~\cite{Gyulassy2007Topologically}, molecular interactions~\cite{gunther2014characterizing}, and astronomy~\cite{sousbie11}.  
A recent survey from Heine et al.\cite{Heine2016survey} lists many other applications of Morse and Morse-Smale Complexes (as well as other topological descriptors).

These recent successes are a direct result of pioneering efforts to develop efficient computations that can extract Morse-Smale complexes~\cite{edelsbrunner03b,gyulassy_vis08} while handling the nuances and scale of scientific data.  
As a result, much of the existing work has focused on visualizing the structures for an individual dataset and presenting the resulting information in a way a user can interpret the wide range of features (see Fig.3 of \cite{GyulassyKKHHP12} for a visual guide).  
While this addresses a common analysis use case, there is relatively less work available on what Morse theory tells us about a \emph{collection} of datasets.  
While there is recent effort to study topological structures of ensembles~\cite{yan2021scalar}, the research community does not yet have a complete picture of how to analyze an ensemble of Morse complexes.  
Notably, unlike with topological summaries such as the persistence diagram and Reeb graph~\cite{bollen2021reeb}, there is no agreed upon definition of distance (or similarity) between two Morse complexes. 
We thus lack the tools, both theoretically and in practice, to visualize a collection of Morse complexes and to discover the possible relationships they contain.

In this work, we utilize techniques from dimensionality reduction to build a feature descriptor of the space of Morse complexes.  
Using this feature space, we can then study a collection of Morse complexes. 
Particularly, we can construct an overview scatterplot of a given collection that helps to identify clusters of related Morse complexes, which in turn indicates clusters of scalar fields that have similar topological behavior.  
Morse theory provides a unique view of topological properties that are intimately coupled with the geometric description, as the complexes one studies are often viewed \emph{directly embedded} in the underlying domain.  
In this work, we seek a feature descriptor which is expressive enough to capture both topological structure as well as their embedding.
Consequentially, we use an image-based representation of the Morse complex, focusing on the boundaries of Morse cells which capture their size and shape. 
This design choice allows us to retain cues of the topological structure (without needing to store the full complex) while also capturing aspects of the geometric embedding.
Conveniently, this encoding is also how we frequently interpret Morse complexes (i.e. through visualization of the arcs), but moreover it also enables a direct usage of image-based tools for dimensionality reduction that are commonly used for machine learning.

  Our approach relies on using autoencoders to first learn a lower dimensional feature space of the image of Morse complexes. Autoencoders have been used as a dimensionality reduction method for many years \cite{kramer1991nonlinear, hinton2006reducing}.  Nevertheless, we find that an autoencoder alone is a suboptimal approach to projecting our dataset into two dimensions.  In a discussion of visualization using t-SNE, the authors suggested that combining the power of an autoencoder that learns a simpler representation of complex data and manifold learning methods like t-SNE may be helpful to address the problem known as the curse of intrinsic dimensionality \cite{van2008tsne}. In this work, we combine the image-based encoding of Morse complexes with the power of autoencoders to extract features from those images.  We then study collections of Morse complexes by projecting the features to 2D using additional dimensionality reduction methods.

    
    We summarize our contribution as the following: we demonstrate an approach that uses an autoencoder to learn a latent representation of an image-based encoding of Morse complexes. The learned representation enables the direct application of dimensionality reduction methods to create a visual space of Morse complexes that capture the decomposition of the domain. We can use such a space to overview a collection of Morse complexes. Using Morse complexes from synthetic functions, microscopy images, and time-varying vorticity fields of flows, we show our proof of concept approach can be used to study the global and local patterns of the collection of Morse complexes.
	
	\section{Related work}

We review related work in the areas of topological analysis, particularly for ensembles, as well as recent works at the intersection of topological data analysis with machine learning.
	
	\subsection{Morse complexes and scalar field analysis}
	
We focus on using the Morse complex as it relates to analyzing scalar fields in the visualization community.  For a theoretical introduction, we refer the reader to Milnor for details on Morse theory~\cite{milnor63}.  
Edelsbrunner and Harer offer a more general reference to the theoretical and computational underpinnings of many related structures~\cite{edelsbrunner2010computational}.


Translating the benefits of the theoretical structure to practical settings presents a number of complexities, particularly for the Morse-Smale complex.  While computing them in 2D was achieved by Bremer et al.~\cite{timo03,timo04} with an approach that subdivided triangles to follow the numerically computed PL gradient, their efficient extraction in three-dimensions has been quite challenging~\cite{edelsbrunner03b, edelsbrunner03}.  Instead, the visualization community has employed approximations based on a discrete encoding of the gradient field~\cite{Defl15, guenther12, gyulassy_vis08, robins11, ShivashankarMN12, ShivashankarN12}, based on pioneering work of Gyulassy et al.~\cite{gyulassy_vis08} to extend Forman's discrete Morse theory~\cite{forman98}.  

From an analysis perspective, one shortcoming of the discrete approaches is the poor geometric reconstruction of gradient flow features, with results strongly biased in the grid directions. To mitigate the biasing problem of steepest-descent, a randomized approach was independently introduced by Reininghaus et al.~\cite{reininghaus2012combinatorial}, for two-dimensional simplicial complexes, and by Gyulassy et al.~\cite{gyulassy_vis12}, for volumetric meshes. This technique produces a better geometric reconstruction and provably converges to the smooth flow features under mesh refinement~\cite{gyulassy_vis12}.	

Variations that focus on particular aspects of the Morse topology have also found use.  Correa et al. construct topological spines to produce a sparse subset of the information encoded in a Morse-Smale complex~\cite{correa11}.  Thomas and Natarajan augment the extremum graph with geometric information for symmetric detect purposes~\cite{thomas2013detecting}.  Gyulassy et al. propose a method to conform a Morse-Smale complex to a given input segmentation, offering a method to incorporate user-driven constraints~\cite{gyulassy_vis14}.  

Regardless of the implementation, the tangible benefits of using Morse complexes, and topological descriptors in general, cannot be understated.  They offer a robust, flexible segmentation approach that can be applied to many different data modalities with mild assumptions.  Coupled with a persistence-based simplification method~\cite{edelsbrunner02}, Morse and Morse-Smale complexes can offer a multi-scale analysis framework that allows for analysis in many different domains.

	\subsection{Ensemble analysis}

Studying a collection of datasets with topological analysis tools presents new challenges for both the scalability and design of visual metaphors.  While there is little work on using Morse complexes, in part because there is no well-defined theory about the extension of Morse complexes to the multivariate setting, there have been applications that utilize different topological descriptors for visualizations of collections of fields.  See Yan et al.~\cite{yan2021scalar} for a recent survey focusing on scalar field comparison.  Due to space limitations, we focus specifically on topology as it extends to multiple fields, and refer the reader to Wang et al.~\cite{wang2018visualization} for a more general survey of ensemble visualization. 

Jacobi sets~\cite{edelsbrunner04} capture the interactions between critical points for multivariate functions, and have shown some promise in feature tracking applications for time-varying events such as molecular interactions and combustion~\cite{bremer2007topological,suthambhara2011simplification}.  Alternatively, extensions of the Reeb graph and contour trees have been used for analyzing multifaceted data.  Joint contour nets offer one such approximation, based on binning the potential ranges of isovalues in the domain space for each facet~\cite{carr14,munch16}.  
H\"{u}ttenberger et al.~\cite{Huettenberger13} use the notion of Pareto optimality to extract extremal structures in multivariate data.
Fiber surfaces~\cite{carr_ev15, klacansky16} offer a more direct generalization of level sets to bivariate data.  
Chattopadhyay et al.~analyzed the projection of the Jacobi set in the Reeb space~\cite{chattopadhyay2014extracting}, introducing the Jacobi structure of a Reeb space.   Tierny and Carr compute a data-optimal extraction of the Reeb space~\cite{tierny_vis16} for studying multivariate fields.  

While the Reeb space and Jacobi sets offer insight into a unified structure of multivariate fields, they are limited to analyzing only small collections at a time (e.g., the dimensionality of the structure is directly related to the number of fields being studied.
Alternatively, recent work has focused studying aggregations of topology through statistical descriptors  One approach is to focus on statistical summaries of data.  Specifically for Morse complexes, Athawale et al. construct a statistical map that characterizes the uncertain behavior of gradient flows~\cite{athawale2020uncertainty}.  Otto et al. model an ensemble and construct pointwise uncertainty in vector field topology, which can be seen as a more general setting of the gradient field behavior captured by Morse theory~\cite{otto10}.  Our work focuses on the structural description of the Morse complex itself, rather than directly considering the gradient flow.

A final approach to studying collections of topological descriptors is through distance metrics. We can use these to compute pairwise distances and study the distribution of features in aggregate.  In particular, there is significant recent work on studying distances for persistence diagrams and graph-based descriptors such as the contour tree and Reeb graph.  These comparative measures offer quantitative indicators that can be incorporated into visualization.  For persistence diagrams, the bottleneck~\cite{cohen2007stability} and Wasserstein~\cite{cohen2010lipschitz} distances have been well-studied.   Computing these distances requires a matching between points, which may be infeasible at scale, and thus authors in visualization have pioneered methods to compute distances progressively~\cite{soler2018lifted} or developed machine-learning based approximations~\cite{qin2021domain}.  

Distances on the persistence diagram are applicable to scalar fields, but might fail to capture some of the more nuanced connective information contained within level sets.  Rieck et al.~augment the persistence diagram with hierarchical information to better capture this~\cite{rieck2017hierarchies}.  More generally, distance measures on level set topology have been defined (see Bollen et al.~\cite{bollen2021reeb} for an extended introduction).  These distances are significantly more difficult to compute directly, as the resulting algorithms require expensive matchings and enumerations.  Sridharamurthy et al.~\cite{sridharamurthy2018edit}, Saikia et al.~\cite{saikia2014extended}, and Beketayev et al.~\cite{beketayev2014measuring} have proposed variants in the simplest case of merge trees that sacrifice theoretical guarantees to produce more tractable computations.  Recently, Yan et al.~\cite{yan2019structural} demonstrated how to construct structural averages of merge trees for use in visualization.

While distance computations offer an appealing component for data analysis, unlike the persistence diagram and Reeb graph, there is no well-studied definition of distance on Morse complexes.  Narayanan et al.~define a distance measure on extremum graphs, offering one possible approach~\cite{narayanan2015distance}.  Another option is to use distances from other domains, such as the F-measure, that are designed to capture similarity in general segmentations.   Gerber et al.~employ this F-measure in the restricted setting of comparing regressed Morse-Smale complexes to their original~\cite{gerber2013morse}.  In lieu of having theoretical work understanding their behavior and properties in data analysis settings, we consider alternatives.  Subsequently, in our work, we take the approach of indirectly capturing relationships through a visual space of Morse complexes rather than relying directly on a quantitative measure of distance for Morse complexes.
	
	\subsection{Deep learning and topology}
	    Deep learning and topological features typically pair in two ways.  First, while there often exist algorithms to compute topological features directly from data, some researchers have created machine learning models to approximate them.  In Reininghaus et al. \cite{reininghaus2015stable}, the authors train a support vector machine to predict persistence diagrams from 3D shape datasets.  Zhang et al. \cite{zhang2018machine} is even more direct; they train a neural network to predict the winding number of a curve (i.e. how many times it encircles the origin).  Other studies, however, focus on using topological features internally.  Hofer et al. \cite{hofer2017deep} created a new kind of neural network layer to embed persistence diagrams into a learned latent space for further computation.  And Hu et al. \cite{hu2021topology} adapted their loss function to weight regions by topological importance, providing better connected segmentation predictions.  Finally, Qin et al.~approximate Wasserstein distance of persistence diagrams using a GAN-based formulation~\cite{qin2021domain}.  While not the most common partners, machine learning and topological data analysis can work together.

\begin{figure*}[h]
    \centering
    \includegraphics[width=0.98\linewidth]{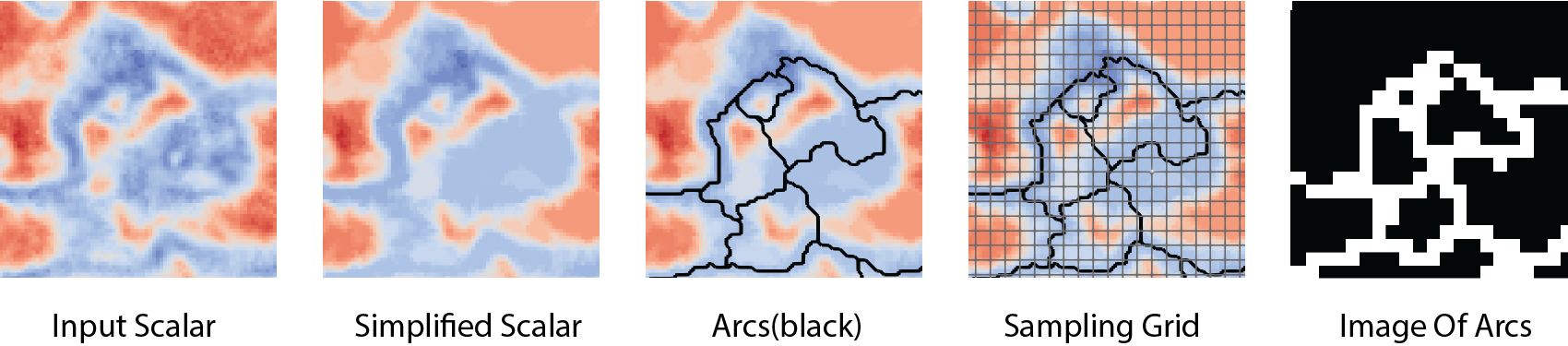}
    \caption{From left to right is our procedure to compute an image of the Morse complex from an input scalar field. We first apply persistence simplification to the scalar field. Then we compute the arcs of its Morse complex (shown as black lines). Finally, for the image resolution of our choice, we label whether a pixel contains at least one arc or not.}
    \label{fig:procedure}
\end{figure*}
	\section{Background on Morse Complexes}
	\label{sec:morse_complex}

We review the key definitions of the topological descriptor we use, the Morse complex, as well as giving and overview of the methods we use to compute it.

\subsection{Definitions}

Let $\domain$ be a smooth manifold, in our setting we consider $\domain$ to be a 2-manifold that is a subset of $\R^2$.  A scalar field $f: \domain \rightarrow \range$ defines a scalar value $f(x) \in \range$ for each $x \in \domain$.  A point $x \in \domain$ is a \emph{critical point} iff $\nabla f(x) = 0$.  A critical point $x$ is \emph{non-degenerate} if the Hessian matrix at $x$ is non-singular.  We say $f$ is a \emph{Morse function} if all of its critical points are non-degenerate and have distinct function values.  In practice, this means that $f$ is well-behaved from a topological standpoint.  Critical points may be further classified by their \emph{index} $\Index$, which on 2-manifolds equals 0 for minima and $2$ for maxima points.  Critical points with indices of 1 are called saddles.


Studying the shape of a scalar field often necessitates understanding the relationships between critical points.  Given a Morse function $f$, an \emph{integral line} is a path on $\domain$ which is everywhere tangential to $\nabla f$.  Topological data analysis can be used to study the relationships between critical points as they are the positions at the limit of paths as one sweeps along integral curves.  
Given a critical point $p$, its \emph{ascending} (resp.~\emph{descending}) \emph{manifold} is defined as the set of points belonging to integral lines whose origin (resp.~destination) is $p$.  The \emph{Morse complex} is the cell complex formed by all descending manifolds.  One can define a symmetric concept for ascending manifolds, which corresponds to the Morse complex of $-f$.

This complex segments $\domain$ into regions associated with critical points.  In 2D, the descending manifolds of maxima are (generically) two-dimensional cells, gathering the region of points $x \in \domain$ that flow uphill and arrive at a particular maxima. Saddle type critical points have a one-dimensional ascending manifold corresponding to the limit behavior on the boundary between two adjacent stable manifolds of two nearby maxima.  These integral lines are called \emph{separatrices}.  For a given two-dimensional cell associated with a maxima, the boundary forms a chain of separatrices connecting alternating pairs of minima and saddles.  These arcs act like a skeleton of the Morse complex, and we rely on a image-based representation of them as a descriptor of the Morse complex itself.



\subsection{Computation}

The robust computation of Morse (and Morse-Smale) complexes has been a long time challenge for the community, as existing algorithms \cite{edelsbrunner03b} were highly complex and required, particularly in 3D, many special cases to account for the transversal intersection condition in 3D.  Fortunately, the alternate formalism of discrete Morse theory enabled a new family of algorithms. 

In discrete Morse theory, we assume the scalar field is represented as a \emph{discrete Morse function} defined over a simplicial complex.  This function assigns a scalar value to every simplex in $\domain$, such that each $i$-simplex $\simplex_{i} \in \domain$ has at most one co-face $\simplex_{i+1}$ (resp. face $\simplex_{i-1}$) with lower (resp. higher) function value: $|\{ \simplex_{i+1} > \simplex_{i} ~ | ~ f(\simplex_{i+1}) \leq f(\simplex_{i}) \}| \leq 1$ and $|\{ \simplex_{i-1} < \simplex_{i} ~ | ~ f(\simplex_{i-1}) \geq f(\simplex_{i}) \}| \leq 1$. Simplices for which these two numbers are zero are called \emph{critical simplices} and their dimension matches their index $\Index$.  

This allows a discrete analogue to $\nabla f$.  A \emph{discrete vector} is a pair of simplices $\{ \simplex_i < \simplex_{i+1}\}$.  A \emph{discrete vector field}  $V$ is a collection of such pairs such that each simplex appears in at most one pair. Then, a \emph{$V$-path} is a sequence of pairs of $V$ $\{ \simplex_i^0 < \simplex_{i+1}^0\}, \{ \simplex^1_i < \simplex^1_{i+1}\}, \dots, \{ \simplex^r_i < \simplex^r_{i+1}\}$ such that  $\simplex_i^j \neq \simplex_i^{j+1} < \simplex_{i+1}^j$ for each $j = 0, \dots, r$. A \emph{discrete gradient} is a discrete vector field for which all $V$-paths are monotonic and loop free. For these, $V$-paths are discrete analogs of integral lines, and simplices which are left unpaired in $V$ are critical.  Separatrices are thus sequences of simplices constructed by extracting ascending (resp. descending) $V$-path(s) from $(d-1)$-saddles (resp. $1$-saddles). 

As we desire an image-based representation for our autoencoder, we construct a binary image of the separatrices of an input Morse complex. Our pipeline is based on the Topology ToolKit~\cite{tierny2017topology}.  First, given an input  scalar field, we (optionally) perform persistence based simplification to preserve selected topological features~\cite{tierny2012Generalized}.  We use this simplification to remove features of persistence less than a threshold.  After simplifying the scalar field, we use TTK to compute the Morse complex. Our goal is to extract the 1-separatrices (arcs) of the Morse complex. 

The arcs are provided as a paths in the original domain. We sample the arcs onto a $n\times n$ resolution image. First the domain is divided into $n\times n$ regular non-overlapping regions. For each region, we assign the label $1$ if it contains at least one arc and label $0$ otherwise. We use those labels for the pixel values of an $n\times n$ image. As the result, we obtain a binary image representing the arcs of the Morse complex. \autoref{fig:procedure} shows an example of the complete procedure to extract image of arcs from a given scalar field.

	\section{Dimensionality Reduction Method}
	\subsection{Autoencoder}
	\label{sec:method_autoencoder}
	An autoencoder is a neural network model that is tasked to reconstruct its own input through sequence of mappings.  
	While often used for self-supervised pretraining \cite{masci2011stacked, laakom2019color}, we apply them to construct a latent space of the input.
	Such spaces are generally compressive, so an $n$-dimensional input will have an $m \ll n$ dimensional latent space \cite{ng2011sparse}.
	An autoencoder usually follows an encoder-decoder type of design. First, the encoder maps the input to a latent representation. Then the decoder tries to reconstruct the input from the latent representation.  Because the latent space is narrow, the network must encode compressed information about the input to accurately reproduce the output from the intermediate representation \cite{hinton2006reducing}.  
	
	Our  goal is to study the latent representation learned by the autoencoder rather than the output of the autoencoder. We place a bottleneck layer in between the encoder and decoder. This bottleneck layer is the output of the encoder. It produces a low-dimensional vector for each image of a Morse complex. We call the vectors \emph{latent vectors}, and the dimensionality of those vectors the \emph{latent dimension}. In our experiments, the latent dimension is one of the main factors that determines the quality of reconstruction. It also impacts what features get preserved in the latent vectors. We investigated the effect of varying latent dimensions in \autoref{sec:results}.
	
	In \autoref{sec:morse_complex} we established our approach to encode the Morse complexes as images. We use convolutional neural networks (CNNs) to set up our autoencoder. We modified commonly used CNNs for image classification including VGG and ResNet \cite{simonyan2014very,he2016deep} as the core of our encoder. In particular, from this set of encoders, we found ResNet-18 performs best in terms of training time and reconstruction quality. ResNet-18 is designed for RGB images of resolution $224 \times 224$. We changed the number of channels (a.k.a. ``inplanes'' in many implementations) so that the input is a single channel gray scale image. The typical design for ResNet-18 reduces the resolution of the input image through residual blocks and pooling layers from $224\times 224$ to $1\times 1$. Since we work with images of resolution $64\times 64$, we removed some pooling layers to compensate for the small input size. We removed the max pooling layer for images of resolution $64\times 64$. For images of resolution $50 \times 50$, we also removed the average pooling layer. The ResNet encoder outputs a 512-dimensional vector. Then we use fully connected layers to map this output to a user specified latent dimension and obtain the latent vector.
	
	Our decoder is a sequence of upsampling and convolutions. We first use fully connected layers to restore the lowest resolution image in the encoder. Then we apply bilinear interpolation to double the images resolution and apply $3\times 3$ convolution to decrease the number of channels of the image. We apply rectified linear unit activation \cite{Nair2010Rectified} after each convolution except the output layer. The output of the decoder is a one channel image of the original resolution, and it is activated by the standard logistic function, $f(x) = \frac{1}{1+e^{-x}}$. So the output the decoder is a gray scale image of values between $(0,1)$.
    
    We jointly optimize the encoder and decoder for image reconstruction. We train the autoencoder to be a function that approximates the identity map of images of Morse complexes. Suppose we have a collection of images of Morse complexes, $\mathbb{I}$.  The autoencoder is then a function $h(x|\theta):\mathbb{I}\to\mathbb{I}$ parameterized by $\theta$, the set of trainable parameters in the autoencoder. Let $f(x) = x$ be the identity map, we train the autoencoder to be an approximation of $f$, i.e. $h(x|\theta)\approx f(x)$.  Note that we can equivalently write $$h(x|\theta) = d(e(x|\theta_e) | \theta_d),$$ where $e(x|\theta_e)=L$ refers to the encoder and parameters (outputting a latent space representation, $L$), and $d(L|\theta_d)$ is the decoder.
    
    This way, we set up the optimization procedure to be $$\argmin_\theta \mathcal{D}(h(x|\theta),f(x))$$ where $\mathcal{D}(a,b)$ is the difference between two images of Morse complexes $a$ and $b$. We consider the difference between two images as the per-pixel difference of the binary classification. So we use a loss function commonly used to measure the quality of binary classification tasks, binary cross entropy loss (BCE loss), as the measure of difference. Given two gray-scale images $a$ and $b$ of size $n$ pixels, whose values are between (0,1) this loss is
    
    $$BCE(a,b) = -\frac{1}{n}\sum_{i}^{n}b_i\log(a_i) + (1-b_i)\log(1-a_i)$$
    
    Therefore, training the autoencoder is the process of finding the best autoencoder parameters $\theta$ to minimize $BCE(h(x|\theta),x)$ for $x$ in the input collection of images of the Morse complex.
    
    In our experiments, we found that ResNet-based autoencoders cannot capture many of the small size features from complex datasets. So in our experiments, we also designed an autoencoder based on a more sophisticated architecture HRNet \cite{wang2020deep}. The HRNet encoder outputs a 512-vector. We use a fully connected layer mapping the $512$-vector to a latent dimension of our choice and then map it back to the $512$-vector for the decoder. The decoder uses a fully connected layer to restore the $512$-vector to $64\times 64$, and pass the $64\times 64$ image through a decoder of HRNet architecture. Later we will show this network design is capable of producing a better reconstruction at a cost of longer training time.
    
	\subsection{Latent space visualization}
	\label{sec:method_latent}
	We consider the autoencoder as a composition of encoding function $e:\mathbb{I}\to\mathbb{R}^n$ which maps an input image to an $n$-dimensional vector and a decoding function $d:\mathbb{R}^n\to \mathbb{I}$ maps that vector back to an binary image, i.e. $h(x) = d\circ e(x)$. We are particularly interested in the output of the encoding function $e$. Given a collection of images of Morse complexes $\mathbb{I}$, we obtain a set of $n$-dimensional vectors $\{e(x)|x\in \mathbb{I}\}$. We study the distribution of $\{e(x)\}$ by projecting those vectors into a 2D space using other dimensionality reduction methods. In our experiments, we tried both PCA and t-SNE. For the t-SNE projection, we use different perplexity values for different datasets. We use ``t-SNE(perplexity)'' to note the parameter we use. We then visualize the collection of Morse complexes by building a scatterplot from the 2D projection.
	
\section{Result and Evaluation \label{sec:results}}
	
    \subsection{Experiments setup and datasets}
    \begin{figure}
        \centering
        \includegraphics[width=0.98\linewidth]{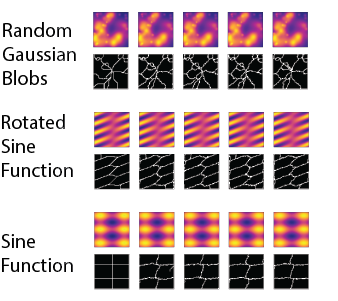}
        \caption{1 sample for each type of scalar function in \textbf{SYNTH}. Each scalar field has 4 noisy variants from left-to-right.}
        \label{fig:synth_sample}
    \end{figure}

    We use 3 different datasets to demonstrate our approach. The first dataset contains images of Morse complexes computed from synthetic scalar fields. We generated each scalar field on a $256\times 256$ sampling grid using helper functions. The first function generates 2 to 512 randomly-placed, symmetric Gaussian ``blobs'' where the standard deviation, $\sigma$, is a uniform random number between 4 and 32. The second function is a 2d sine function defined as $f(x,y) = \sin(\frac{x}{\alpha}) + \sin(\frac{y}{\beta})$ where $\alpha$ is a random integer in $[5, 20]$ and $\beta$ is a random integer in  $[10, 40]$. We also considered a rotated orientation by adding two rotation terms so the the function becomes $f(x,y) = \sin(\frac{x}{\alpha} + \frac{y}{\gamma}) + \sin(\frac{y}{\beta} + \frac{x}{\delta})$ where $\gamma$ and $\delta$ are random integers in $[10,80]$ with 50\% probability of being negative. 
	
	For each scalar function we generated, we also generated 4 scalar fields with uniformly random noise of magnitude $[0,0.05]$ added to the base function to generate 4 additional variants. Then we applied persistence simplification removing all features with persistence less than $0.04$ from all 5 scalar fields.  This process creates fields that have similar topology structures, but the noise perturbation can cause subtle shifts in the arc geometry.
	
	Finally, Morse complexes are computed on the base $256\times 256$ grid, and the image of the Morse complex is sampled to $64\times 64$ images. We generated a total of 10,000 images. In the end, we obtain a set of Morse complexes from 3320 random Gaussian blobs, 3435 sine functions without rotation, and 3245 sine functions with rotation. \autoref{fig:synth_sample} shows 1 sample for each type of function in our dataset, the image of arcs are generated from persistence simplified scalar fields. We refer to the images of arcs extracted from this dataset as \textbf{SYNTH}.
	
	\begin{figure}
	    \centering
	    \includegraphics[width=0.98\linewidth]{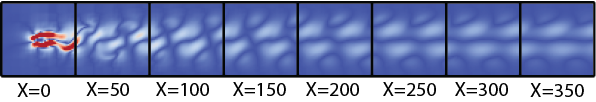}
	    \caption{Our sampling strategy at each time step of \textbf{FLOW}. We sample 8 disjoint $50\times 50$ subsets from the $400\times 50$ scalar fields.}
	    \label{fig:c2d_sampling_strategy}
	\end{figure}
	The second dataset comes from 2D flow over cylinder simulation by Tino Weinkauf \cite{weinkauf10c} using Gerris flow solver\cite{gerrisflowsolver}. We computed the vorticity magnitude from the $u,v$ vector field using the VTK Gradient filter\cite{VTK4}. The scalar fields were sampled on a $400\times50$ regular grid and from $1001$ timestamps. We extract $50\time 50$ crops translating across different $x$ coordinates. These crops cover the whole domain, with the leftmost $x$-value ranging from $0$ to $350$ with the spacing of $50$. So for each $400\times 50$ scalar field, we produce 8 disjoint $50\times 50$ subsets. We extract Morse complexes from those $8008$ subsets to obtain $8008$ $50 \times 50$ images of arcs. We did not apply any persistence simplification. We refer to this dataset as \textbf{FLOW}.
	
	The third dataset comes from MICCAI Challenge on Circuit Reconstruction from Electron Microscopy Images (CREMI)\footnote{\url{http://www.cremi.org}}. There are 3 pairs of volumes labeled as $A$, $B$, and $C$ taken from different parts of adult fly brain tissues. Volumes encode data using greyscale values in $[0,255]$.  Each pair contains two volumes for training and testing. Each volume has resolution of $1250\times 1250\times 125$. Where the distance in the $z$ direction is $10$ times large than those for $x$ and $y$ directions. 
	
	We use only the $A, B$ volumes for training as our example datasets. We randomly selected 30000 positions from those volumes and took $64\times 64$ slices in $xy$-plane. The Morse complexes are extracted from those slices with a persistence simplification threshold of $30$. We then sampled the Morse complexes to $64\times 64$ binary images. We sample another 6,000 Morse complexes from those not seen in the training dataset as the testing dataset to ensure the trained model generalizes to the unseen parts of the volumes. We refer to this dataset as \textbf{CREMI64}.
	
	From the same pair of volumes, we randomly selected 30,000 $512 \times 512$ slices. We extract Morse complexes with a persistence simplification threshold of $50$. We then sampled the arcs to $64\times 64$ binary images as a training dataset. We again extracted 6,000 Morse complexes from volumes unseen in the training set and prepared them as the testing dataset. We refer to this dataset as \textbf{CREMI512}.
	
	\subsection{Autoencoder training and reconstruction quality}
	
	\begin{figure*}
        \centering
        \includegraphics[width=0.98\linewidth]{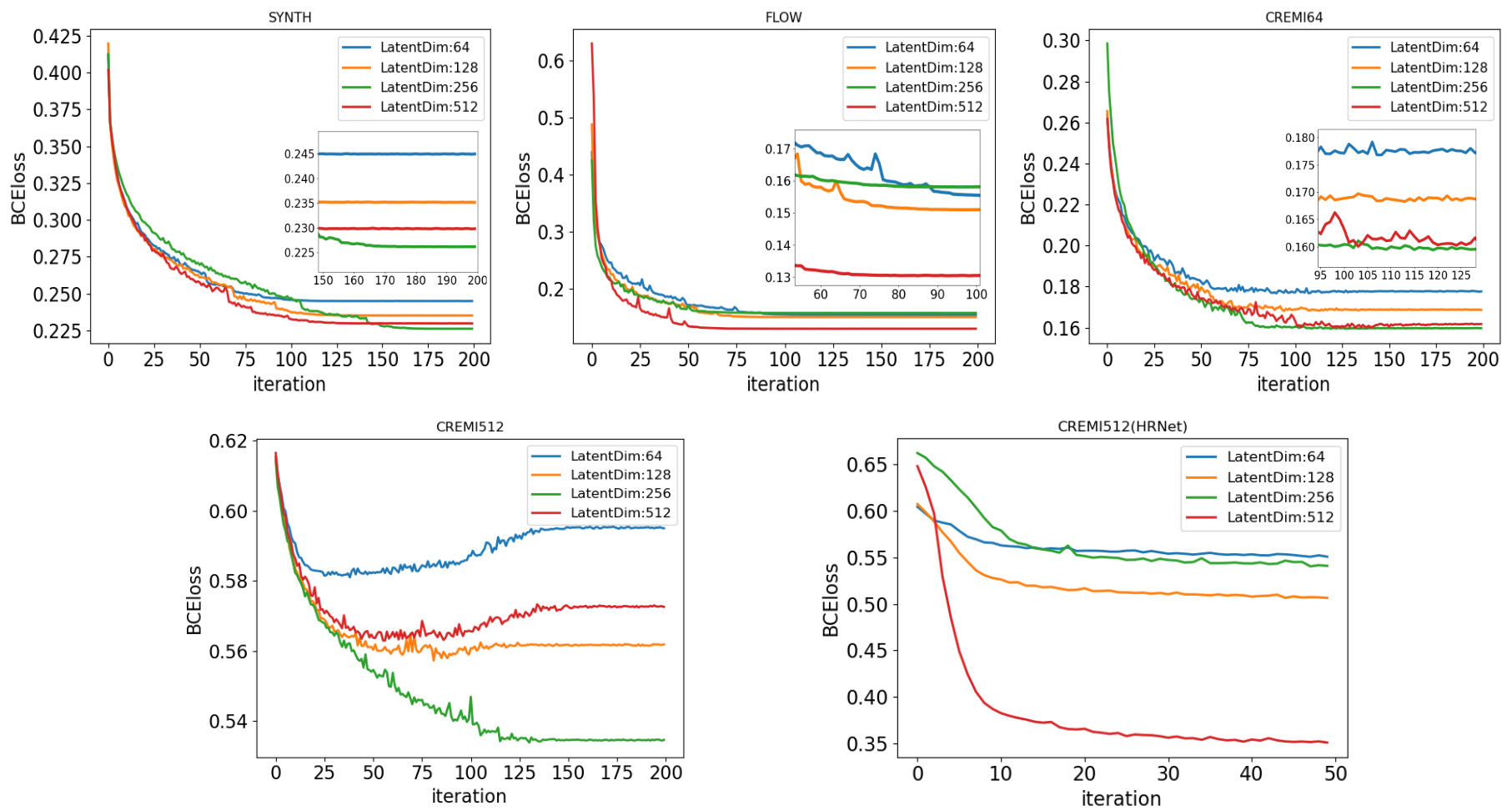}
        \caption{Reconstruction loss (lower is better) vs.~iteration for ResNet-based autoencoders with varying latent dimension, trained for all 4 datasets. Bottom right: the HRNet-based autoencoder trained on \textbf{CREMI512}.  These plots help show the best combination of encoder and latent dimension.}
        \label{fig:quality_vs_iterations}
    \end{figure*}
	
	For each dataset, \textbf{SYNTH}, \textbf{FLOW}, \textbf{CREMI64}, and \textbf{CREMI512}, we trained several ResNet-18 based autoencoders with different latent dimensions in $\{64,128,256,512\}$. For \textbf{CREMI512} we also trained HRNet-based autoencoders. We implemented autoencoders using PyTorch. For all the training in this paper, we used the Adam optimizer\cite{kingma2014adam} with an initial learning rate of $1e-4$. The learning rate halves whenever the training loss reaches a plateau, i.e.~when the training loss stops dropping compare to the previous iteration. For all the ResNet autoencoders, we train them for 200 iterations. For the HRNet autoencoders, we train for 50 iterations. The training was done on an Intel(R) Core(TM) i7-7700K CPU @ 4.2 GHz with an NVIDIA GeForce 1080 Ti GPU with CUDA 11.4. 
	
    Although we do not use the output of the decoder (the reconstruction of the input) in this project, we want the reconstruction to be reasonable for validation. If the decoder can recover a good reconstruction from the output of the encoder, then we are more confident that the latent vectors contain at least enough information to describe the input Morse complexes.
    
    There are several factors that may impact the reconstruction quality of the autoencoder, from the network architecture and other hyper parameters to the random seeds we use for each model. The optimal autoencoder configuration depends on the distribution of input data. For this work, we want to obtain a latent space that can help us visualize the collection of Morse complex, so we primarily investigate the parameter that defines the latent space, the latent dimension. 
    Another important factor that might impact the reconstruction quality is the random initialization of the autoencoder's weights. In our experiments, we tried 50 random seeds from 0 to 49 for each model and chose the best we encountered to report in this section.
    
    \subsubsection{Quantitative results for choosing the best model} 
    
    For all the autoencoders we trained, we plot the BCE loss of each autoencoder against the number of iterations (sometimes called ``epochs'') they are trained. The results are shown in \autoref{fig:quality_vs_iterations}. We trained all the ResNet-based models for 200 iterations. After a certain iteration, the gain for training is limited allowing for early termination. We want to choose the autoencoder with the best reconstruction quality (lowest BCE loss) for our latent space visualization for each dataset. 
    
    For the \textbf{SYNTH} dataset, the BCE loss is computed on all samples of the training dataset. After the 150th iteration, the autoencoder with 256 latent dimensions performs the best. For this dataset, we use the autoencoder trained for full 200 iterations for later tasks. The training takes a total of \textbf{46 minutes}.
    
    For \textbf{FLOW} dataset, the BCE loss is computed on all samples of the training dataset. we see the autoencoder with 512-dimensional latent space outperforms the other three. This is surprising because this dataset is the simplest --- suggesting that a smaller latent dimension would be sufficient. The average BCE loss per pixel is just a little above 0.13: better than 0.225 for \textbf{SYNTH} and 0.16 for \textbf{CREMI64}. Unlike the other two datasets whose autoencoders with latent dimension 256 outperform others, the autoencoder with latent dimension 256 for \textbf{FLOW} dataset performs worst. We believe there are two possible factors here. First, the random seed for weight initialization does change the quality of prediction, so they may change the ranks of reconstruction performance. The result shown here is the best from multiple trials, but it is not guaranteed to be the absolute best. Second, the image of arcs for \textbf{FLOW} dataset are not resampled, so they only have a resolution of $50\times 50$, compared to $64\times 64$ for other datasets.  To compensate for the ResNet architecture, we removed the average pooling layer before the fully connected layers. This architecture change could explain how reconstruction quality responds to changes in the latent dimension. For this dataset, we chose the autoencoder of latent dimension 512 trained for 80 iterations to use for later tasks.  Training this autoencoder for 80 iterations takes \textbf{11 minutes}.
    
    \begin{figure}[!ht]
        \centering
        \includegraphics[width=0.98\linewidth]{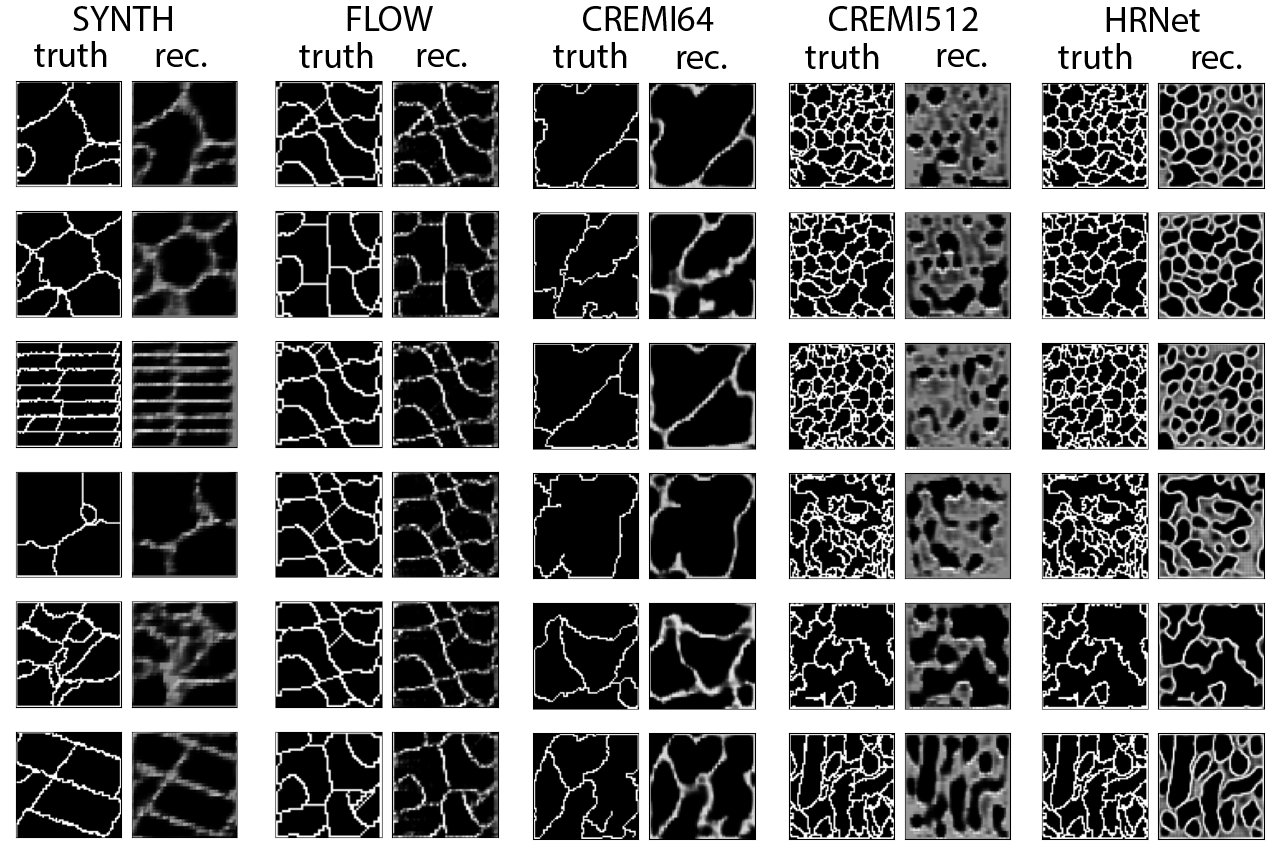}
        \caption{Reconstruction quality of different network models. The results of \textbf{SYNTH} are from the ResNet autoencoder with latent dimension 256 at the 200th iteration. The results of \textbf{FLOW} are from the ResNet autoencoder with latent dimension 512 at the 80th iteration. The results of \textbf{CREMI64} are from the ResNet autoencoder with latent dimension 256 at the 120th iteration. The results of \textbf{CREMI512} are from the ResNet autoencoder with latent dimension 256 at the 150th iteration. The rightmost group of results is from the HRNet-based autoencoder with latent dimension 512 at the 50th iteration.}
        \label{fig:visual_quality}
    \end{figure}
     
    For \textbf{CREMI64} dataset, the autoencoders are trained on the training dataset with $30,000$ samples, losses are evaluated using $6,000$ samples from the testing dataset. The autoencoder with latent dimension 256 outperforms the others. The reconstruction quality of that autoencoder only marginally improves after the 100th iteration. We observe that in the 100th to 105th iterations, the latent dimension 512 happens to outperform the others. We choose, however, the autoencoder with 256 latent dimensions trained for 120 iterations for this dataset.  Training this autoencoder for 120 iterations takes \textbf{60 minutes}.
    
    For \textbf{CREMI512} dataset, we use the $6,000$ samples from the testing dataset to plot the loss. The qualities of reconstruction are poor across all ResNet-based autoencoders. The best BCE loss per pixel is higher than 0.53 compared to other ResNet models for other datasets that have 0.225 at most. For all autoencoders other than the 256-dimensional latent space one, they show signs of overfitting after too many iterations. Overfitting in this context means the autoencoder is trained for optimizing the training dataset, such that the resulting autoencoder performs poorly on the testing dataset (i.e.~it fails to generalize). It is not shown in the plot, but the training loss monotonically decreases for all autoencoders for all 200 iterations (whereas the testing loss as shown fluctuates). So the increase in testing loss is an indicator that overfitting occurs in the training processes. The best autoencoder to use for this dataset is the 256-dimensional latent space one trained for 150 iterations. Despite the reconstruction quality, we observe that this autoencoder can capture large size features and produce interesting latent spaces. Training this model for 150 iterations takes \textbf{76 minutes}.
    
    Since the ResNet-based autoencoders for complex datasets like \textbf{CREMI512} perform poorly, we experimented with a much more complex model which is based on HRNet as we mentioned in \autoref{sec:method_autoencoder}. We only trained those networks for 50 iterations. It takes \textbf{9 hours} to train 50 iterations of the HRNet-based autoencoder on 30,000 $64\times64$ training images. For each iteration, we also evaluate the model using 6,000 testing images. The error plot shows that the best model is the one with a 512-dimensional latent space.

    \subsubsection{Qualitative results of the reconstruction} 
    
    Next, we will show the reconstruction results from the models we chose. \autoref{fig:visual_quality} shows the ground truth (input of the autoencoders) and the reconstruction result side-by-side. The results labeled \textbf{CREMI512} and HRNet are showing the same set of samples from \textbf{CREMI512} dataset. They are produced using ResNet-based autoencoder and HRNet-based autoencoder respectively.
    
    \begin{figure}[!ht]
        \centering
        \includegraphics[width=0.98\linewidth]{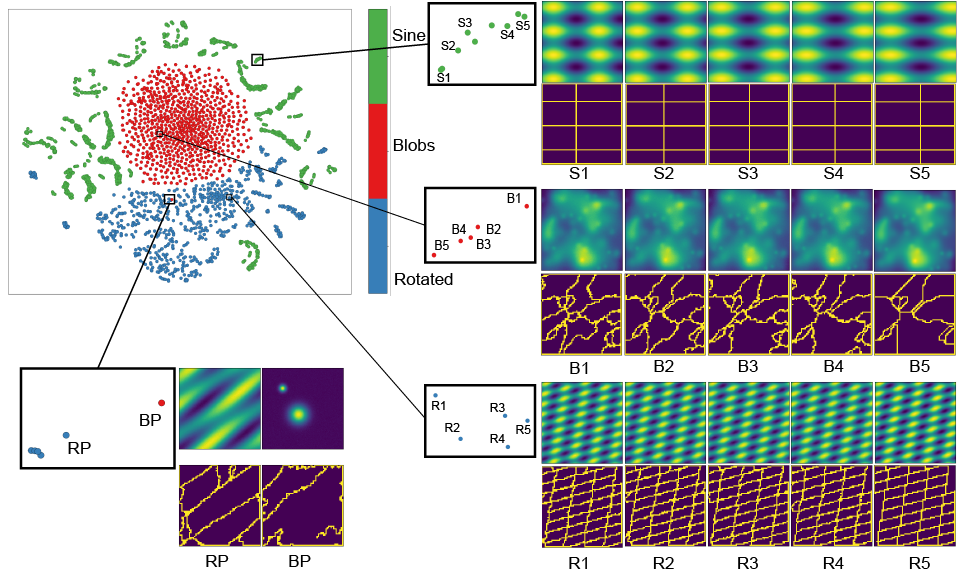}
        \caption{Scatterplot of 10,000 Images of Morse complexes projected using t-SNE(30) from the synthetic dataset. Points are colored by which type of function they come from.}
        \label{fig:synth_overview}
    \end{figure}
    
    For \textbf{SYNTH}, \textbf{FLOW}, and \textbf{CREMI64}, the ResNet-based autoencoder reconstructions are blurred version of the original input. Same as the quantitative results, the reconstructions for \textbf{FLOW} dataset are the best among all three. The gray-scale reconstructions mostly resemble the shape of the ground truth. In many places, it is even capable of capturing small size features that other ResNet-based autoencoders could not. The \textbf{CREMI64} autoencoder is slightly worse than the \textbf{FLOW} autoencoder, especially in the regions near the crossing of multiple arcs.  In such places, the \textbf{CREMI64} autoencoder tends to produce a region of gray. For some small features, such as the bottom of the second row, the reconstruction acknowledges there are arcs in that region, but it fails to capture the shape of the arcs. For \textbf{SYNTH} dataset, we notice that it performs better for more regular shapes like the sine functions and rotated sine functions. The random Gaussian function, especially in the region where multiple small features are placed together, tends to produce a region of gray pixels. These three autoencoders capture the position and overall shape of the large region of black (corresponding to large Morse cells).
    
    For \textbf{CREMI512} dataset, the ResNet-based autoencoder failed to reconstruct any regions with densely distributed white pixels. The black regions in the reconstructed images often correspond to large features in the ground truth. But as the top half of the 4th-row shows, this model sometimes fails to capture large regions of black as well. Even for simpler input like the one shown on the 5th row, this ResNet-based encoder produces a poor reconstruction. On the other hand, the HRNet-based autoencoder is capable of capturing more features. Compared to the ResNet-based autoencoder, the number of features that are visible in the HRNet reconstruction increased. And we see the visible features in the reconstruction better capture their true shapes. This autoencoder also suffers, however, from bad quality of smaller features. But it appears to be an overall improvement to the ResNet-based autoencoder, even if it comes at a significant computational cost. 
    

    \subsection{Latent space}
    \subsubsection{Synthetic functions}

    \begin{figure}[!ht]
        \centering
        \includegraphics[width=0.98\linewidth]{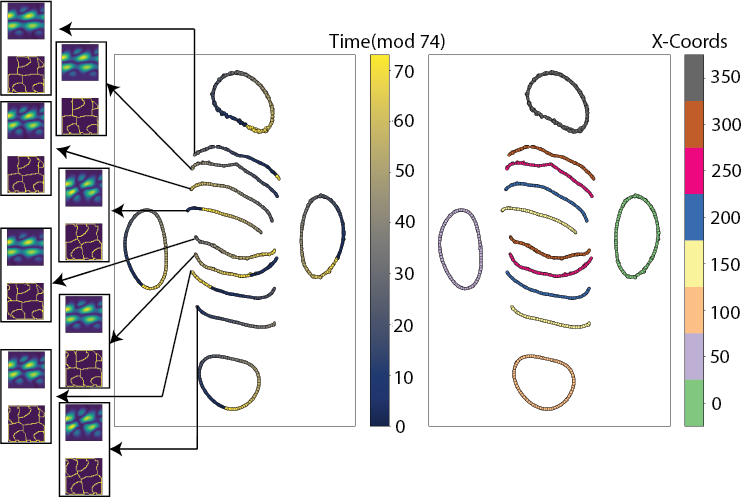}
        \caption{Scatterplots show the t-SNE projection of 8008 Samples from \textbf{FLOW} dataset. We first project the samples to 256-dimensional latent space using the ResNet-based autoencoder trained for this dataset. And use t-SNE(60) to project the samples to 2D. The left scatterplot is colored by the time-stamp of the samples $(\mod 74)$. The right scatterplot is colored by the samples' $x$-coordinate.}
        \label{fig:c2d_overview}
    \end{figure}
    
    \autoref{fig:synth_overview} shows a scatterplot of 10,000 images of arcs from 3 types of synthetic functions. The feature vectors are obtained from the latent space of the ResNet-based autoencoder with 256 latent dimensions. The 2D scatterplot is generated by projecting latent vectors using t-SNE(30). The points are colored by their function types (blob, sine, or rotated sine). Overall, the same color points are projected together. The sine function images without a rotation factor are scattered around the outer circle of the plot. Sine functions with rotation concentrate in the bottom half of the plot. And Gaussian blobs concentrate in the  upper-middle region. There is one particular image from Gaussian blobs (labeled BP in the figure) that is placed in the neighborhood of rotated sine functions. With a closer inspection, we found that the nearest neighbor of BP in the scatterplot does show similar shape features as BP. Their arcs are tilted in a similar direction, and the largest feature in BP looks like the two largest features in RP merged together. This suggests for this synthetic data set that our projection is able to group members according to the geometry of their Morse complexes.
    
    Zooming into the scatterplot we found that many of those points are grouped into smaller clusters of 5 points. Typically, those 5 points are from the same base (with no random noise) function. But, the sine function without rotation factors shows a different type of clustering behavior. From the points we selected, we examined this particular cluster and found all the points from this cluster are sine functions without random noise. And their noisy variants are projected in a different nearby cluster. For those functions, the points distribution follows the $1+4$ pattern, meaning the 4 noisy variants are placed together. This also suggests in our sampling strategy, we are much more likely to sample similar sine functions without rotation. For the other two types of synthetic data sets, their base function images are grouped with noisy variants. This suggests that, in terms of image difference, the similarity between noisy and non-noisy Morse complex is higher than between two samples of different parameters.

    \subsubsection{2D flow over cylinder}

    The flow dataset contains periodic behavior at a period of 74. For the left plot of \autoref{fig:c2d_overview}, we color the scatterplot using timestamp $t (\mod 74)$. For the right plot, we color the points using each sample's $x$-coordinate. From the combination of those two plots, loop-like structures emerge within the latent space. This particular instance is selected from multiple trials of t-SNE(60). The appearance of these loop structures, however, depends on t-SNE initial conditions.  What is consistent between trials is that the samples from earlier $x$-coordinates (near the source of the flow), especially those from $x=0,50,100$ are most likely to be projected as loops in the projection. They rarely are broken into two segments. The middle and later samples, especially $x=200,250,300$, never form a loop. Sometimes the last sample at 350 forms a loop.
    
    In this particular instance of the projection, There are 4 loops in the projection for samples from $x=0,50,100,350$. The rest of the samples are broken into two segments of sequence. As we examine the point where the sequence break in the projection, we found that all the Morse complexes on the top left endpoints do not have a Morse cell on their top left, whereas Morse complexes on the bottom left endpoints have some feature there. Apart from that difference, the rest of the Morse complex looks very similar in terms of the shape and position of the features. And not only the endpoints for the same $x$ positions are similar, across different $x$-coordinates, but the Morse complexes for all left endpoints are also similar in terms of feature geometry. For Morse complexes at the right endpoints of the segments, we observe similar behaviors.
    
    For this dataset, the latent space shows patterns matching our assumption of the dataset. First, it shows the sequential pattern of the Morse complexes. We see the samples in the latent space are organized into sequences according to the timestamp. Then we can conclude the Morse complex of samples also shows the sequential pattern we expected from the scalar field. They also form loops that align with the periodicity of the scalar fields. Second, it groups the samples by the $x$-coordinates. This suggests, in our sampling strategy, that the difference in the topology between different $x$-coordinates is more noticeable than the difference in the temporal domain from the latent space. We believe this is the result of discretely sampling the spatial domain while having relatively dense sampling in the temporal domain.

    \begin{figure}
        \centering
        \includegraphics[width=0.98\linewidth]{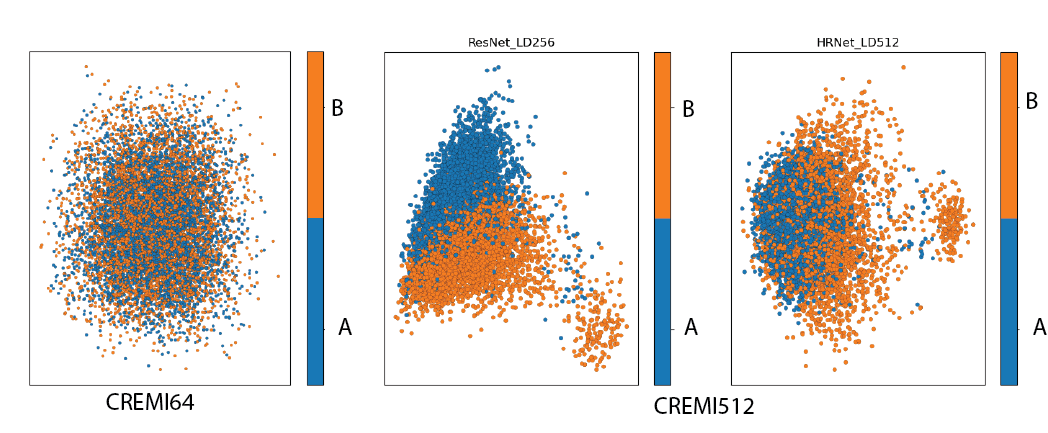}
        \caption{The plot in the left shows 10,000 random samples selected from \textbf{CREMI64} dataset projected using autoencoder with 256-dimensional latent space. The plots in the middle and right show 10,000 random samples from \textbf{CREMI512} dataset. The middle plot projects the samples using the ResNet-based autoencoder. The right plot projects the samples using the HRNet-based autoencoder. All the latent spaces are projected to 2D using PCA. Especially for \textbf{CREMI512} dataset, we found PCA projections provide better separation between samples from $A$ and $B$ than all the t-SNE projections.}
        \label{fig:cremi_latent}
    \end{figure}
    
    \subsubsection{CREMI}

    Unlike the last two, we made fewer assumptions about this dataset. The Morse complexes are computed on the slices from two different volumes, $A$ and $B$, for which little distinguishing characteristics are provided. We used our approach to investigate the difference between the Morse complexes between the two volumes. \autoref{fig:cremi_latent} shows three scatterplots of the PCA projections of different latent spaces. 
    
    We chose PCA over t-SNE as the projection method for all samples of CREMI data because PCA projections for this dataset provide better separation between samples from $A$ and $B$ for \textbf{CREMI512} dataset. We experimented with several runs using t-SNE(15), t-SNE(30), t-SNE(50) (not shown) on both \textbf{CREMI64} and \textbf{CREMI512} datasets. For all the t-SNE projections, we observe no separation between samples $A$ and $B$.  PCA projection provides some structure in how Morse complexes are distributed in the visual space for \textbf{CREMI512} dataset.
    
    The leftmost scatterplot shows the PCA projection of $10,000$ samples from \textbf{CREMI64} dataset. The scatterplot shows no separation between $A$ and $B$. From examining the plot and the original dataset, we conclude that at the resolution of $64\times 64$, only a few Morse cells are captured in each slice, resulting in only limited topological information.  The Morse complex differences between slices from $A$ and $B$ are not apparent in this resolution. 
    
    This motivated enlarging the resolution of scalar field slices to $512\time 512$ to build the dataset \textbf{CREMI512}. The two scatterplots on the right show the PCA projected latent spaces of the ResNet-based autoencoder and HRNet-based autoencoders of 10,000 random samples from \textbf{CREMI512} dataset. Along the major axis of the PCA projections ($x$-axis of the scatterplot), the Morse complexes' complexity decreases for both latent spaces. More precisely, on the left side of both projections, the Morse complexes tend to have a large amount of small Morse cells. On the right side, the Morse complexes tend to have a few large Morse cells. Therefore, for both projections, the latent spaces show that it is more likely for samples from $B$ to have few large Morse cells. In terms of scalar fields, those large Morse cells correspond to large cellular structures in the images.
    
    From both latent spaces, we can see some separation between samples from $A$ and $B$. For the ResNet latent space, the samples from $A$ mostly appear on the top half of the scatterplot. The samples from $B$ appear mostly in the bottom half of the plot. In this latent space, there is a region on the top of the plot where all the samples are from $A$. But in the HRNet latent space, all samples from $A$ are mixed with samples from $B$. There are no groups of samples from $A$ separate from samples from $B$.
    
    \begin{figure}
        \centering
        \includegraphics[width=0.98\linewidth]{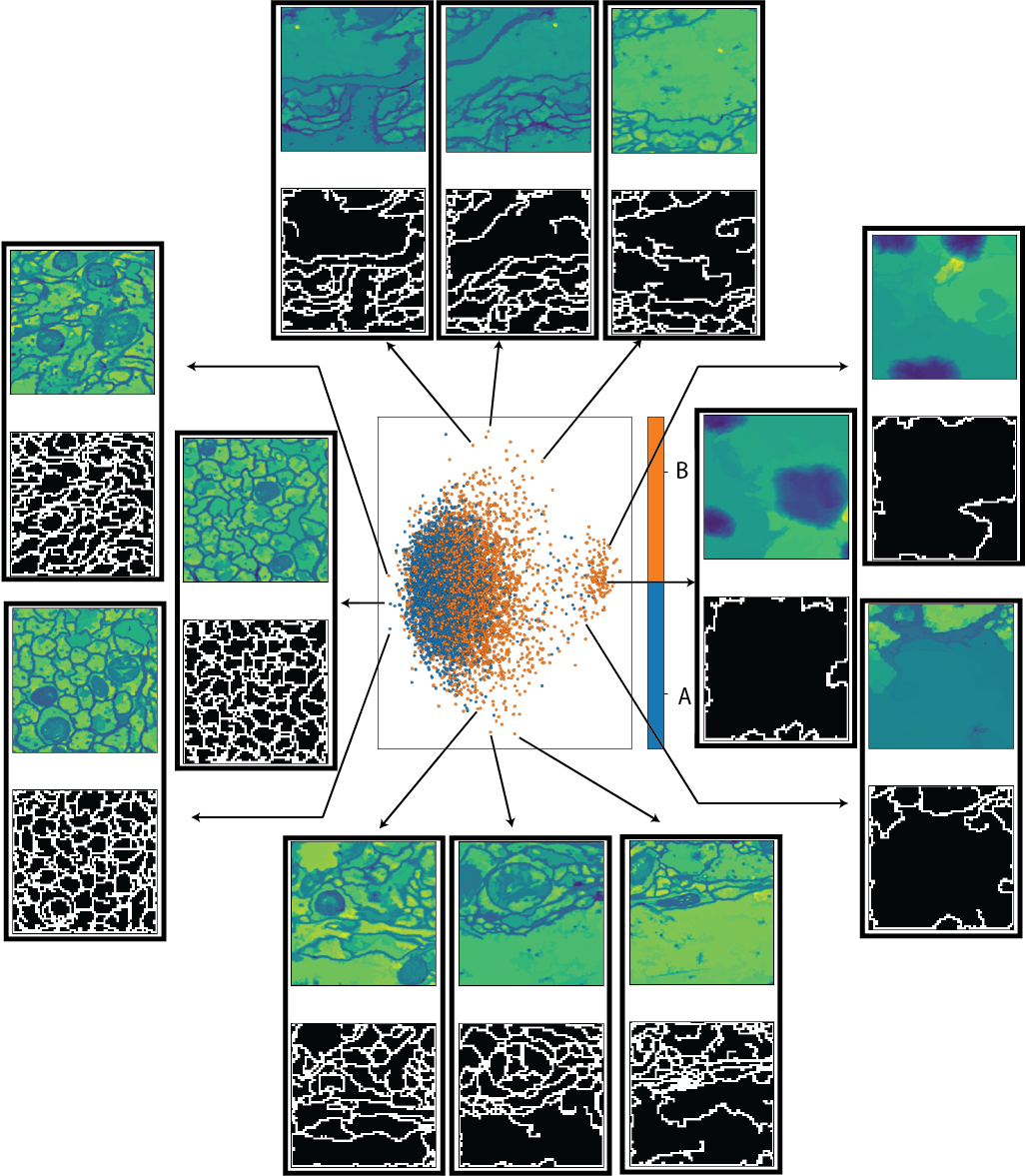}
        \caption{9 example scalar field and Morse complex pairs selected from different locations in the PCA projection of the HRNet 512 dimensional latent space of \textbf{CREMI512} dataset. The scatterplot shows all 10,000 samples. The latent space is structured in the following way: From left to right, the Morse complexes decrease the complexity. On the top of the scatterplot, the Morse complexes have large Morse cells (corresponding to cellular structures in microscopy images) in their top half, while at the bottom we see this trend reversed.}
        \label{fig:HRNet_latent}
    \end{figure}
    The HRNet latent space organizes those 10,000 samples in a more structured way (see \autoref{fig:HRNet_latent}). On the left side of the plot where $B$ and $A$ are mixed together, the samples tend to have a lot of small Morse cells. Moving right, the number of samples from $A$ decreases, and larger cells start to appear in the samples. In the middle, we can see there are a few samples from $A$ scatters, but there are a lot more samples from $B$. On the middle-top region of the scatterplot, the samples from $B$ have large Morse cells on the top half of the Morse complex. On the middle-bottom region of the scatterplot, the samples from $B$ have large Morse cells in the bottom half of the Morse complex. In our experiment, since we randomly select 10,000 samples from \textbf{CREMI512}, the PCA placement in the $y$-direction may be flipped in some trials. So a large Morse cell at the top of the Morse complex may be placed at the bottom of the projection. But in the positive $x$-direction, the complexity of Morse complexes consistently decreased in all our trials.
    
    Overall, both ResNet- and HRNet-based autoencoders agree that samples from $B$ are more likely to have large features in them. And this key difference between samples from $A$ and $B$ was not identified from samples using low-resolution scalar fields. 

	\section{Discussion}
	
	
	
	
	
	We demonstrate an approach to analyzing a collection of Morse complexes that combines image-based encoding of Morse complexes, autoencoders as feature extractors, and dimensionality reduction methods to map the collection into a visual space. While promising, there are numerous opportunities for further investigation.  We focused mainly on ResNet-based autoencoders, but note that the field of machine learning is rapidly evolving, and there are numerous additional possibilities to consider. Rather than developing the best possible architecture, we emphasized proof-of-concept and feasibility.  We note that there are many options for including additional tools for regularization on latent vectors, such as sparse autoencoder\cite{ranzato2006efficient}, variational autoencoders\cite{kingma2013auto} or adversarial autoencoders\cite{makhzani2015adversarial}, all of which might generate more interesting latent spaces. Interestingly, we found that an autoencoder alone offered less utility than combining an autoencoder with dimensionality reduction techniques like PCA and t-SNE.  Likewise, using only the dimensionality reduction approach was suboptimal (in addition to being time-consuming).  This echoes observations of other authors, for example, Gadirov et al.~who consider dimensionality reduction of spatial data~\cite{gadirov2021evaluation}.

    On the visualization side, we used 2D scatterplots to display the overview. They help to demonstrate the overall structure of the data. However, the local neighborhoods and details of individual Morse complexes need to be inspected through interactions. We believe automatically highlighting representative samples would help understand data collection. Since we compute the latent vectors in a data-dependent way, and the 2D point distribution also depends on the projection method, automatically identifying representative points from the scatterplots could be very difficult. In our results, we manually selected samples to inspect the details.
	
	On the topological-feature side, we intentionally chose an image-based representation of arcs, as this captures both a common feature \emph{and} its embedding, but note there are other possibilities that come with different challenges.  One could consider a similar framework that uses Morse cell labels as segmentation ids, but this would require a different model for the autoencoder, particularly since it would need to accommodate mapping two Morse complexes with similar shapes (but different labels) into a similar feature.  Alternatively, one could consider taking the Morse complex as a cell complex and utilizing graph neural networks\cite{wu2020comprehensive}.  This approach might better capture the connectivity of cells, but it would create additional challenges in encoding the arc geometry within the input data, as an abstract graph representation alone would discard it.  Scalability is also a concern, as we focused primarily on 2D datasets, for which there are many readily-available neural networks, rather than tackling 3D in this work.  Finally, our results on \textbf{CREMI64} indicate that to utilize autoencoders, a careful balance between feature scale and image resolution must be maintained.  How best to achieve this in a more automated fashion requires additional future work.

\acknowledgments{We thank the anonymous reviewers for their valuable opinions and comments. This work is supported in part by the U.S. Department of Energy, Office of Science, Office of Advanced Scientific Computing Research, under Award Number(s) DE-SC-0019039.}

\bibliographystyle{abbrv-doi}

\bibliography{main,josh}
\end{document}